\documentclass[aps,preprint,preprintnumbers,nofootinbib,showpacs]{revtex4}
\usepackage{graphicx,color,amsmath}

\begin{document}
\title{Direct CP violation in $B^\pm\to  p \bar p K^{(*)\pm}$}
\author{C. Q. Geng$^{1,2}$,
Y. K. Hsiao$^{1,2}$ and J. N. Ng$^{2}$}
\affiliation{$^{1}$Department of Physics, National Tsing Hua University, Hsinchu, Taiwan 300\\
$^{2}$Theory group, TRIUMF, 4004 Wesbrook Mall, Vancouver, B.C. V6T 2A3, Canada}
\date{\today}
\begin{abstract}
We study the direct CP violation in $B^\pm\to p\bar p K^{(*)\pm}$
decays in the standard model.
 We point out that these three-body baryonic $B$ decays can
be important tools for detecting the
direct CP violation in the charged $B$ system,
in which there are no conclusive signatures yet.
In particular, we show that
the direct CP violating asymmetry in
$B^\pm\to p\bar p K^{*\pm}$ is around 22$\%$ which supports the
recent data by the BABAR Collaboration.
\end{abstract}

 \maketitle
\newpage
One of the most important purpose to study $B$ physics is to
test the mechanism of CP violation
as given in the
Cabbibo-Kobayashi-Maskawa (CKM) framework \cite{CKM} of the
Standard Model (SM)
and search for new physics. Although the data of the
mixing induced CP
asymmetry in $\bar B^0\to J/\Psi K_S$ \cite{JPsiAcpBABAR,JPsiAcpBELLE} as well as
the direct CP asymmetry in $\bar B^0\to \pi^+ K^-$ \cite{KpiAcpBABAR,KpiAcpBELLE}
and $\bar B^0\to\pi^+\pi^-$ \cite{pipiAcpBELLE} have been
 measured precisely, analogous observations for the charged modes are not conclusive yet.
The situation is similar even in $K^{\pm}$ decays. To study CP violations
in charged $B$ decays using two body modes  there exist challenges  in both theory and
experiments.
For example, these
charged modes involve the neutral final states which make for low
experimental efficiencies. Moreover, the present data fail
to match the expectations of  the SM in the $B^{\pm}\to K^{\pm}\pi$ decays.
The SM predicts that  some of them
should have a direct CP asymmetry  very close to those of their
neutral partners \cite{KpiAcpBELLE,charged}. As a consequence, there appears to have  room for the
contribution of new physics \cite{Buras1,Buras2,Botella}.
On the other hand,
even some CP
violations are found in the promising modes, such as
$B^{\pm}\to\pi^{\pm}\pi^0$
and $B^{\pm}\to\pi^0 K^{\pm}$, clear theoretical understandings
would  still be hard to achieve  due to the hadronic uncertainties
\cite{Ali:2004dq,Brev,Gronau:2006xu,Lipkin:2006pc,ICHEP06-1,ICHEP06-2}.
It is
therefore crucial to thoroughly examine CP violation in the charged $B$
system
because not only it is a long standing puzzle but also
it may hide new physics.

In this letter, we will study the direct CP violation in
the three-body baryonic charged $B$ decays of
$B^\pm\to p\bar p K^{(*)\pm}$. Note that there is no neutral final state
in the decays and hopefully that will make for a clean experimental signature at the
price of lower branching ratio.
Recently, the
 experiments \cite{Wang1,angularppK,CPppK} with threshold
effects \cite{Threshold} and large forward-backward angular and
Dalitz plot distribution
asymmetries have eliminated large uncertainties from the
strong interactions as they reveal more information for the
decay mechanisms than the two-body ones.
As a result, the decays of $B^\pm\to p\bar p K^{(*)\pm}$ could provide
good probes to investigate the weak phase in the
charged $B$ system. We shall demonstrate that these three-body baryonic
$B$ decays can be important tools for CP violation due to their
simple amplitudes theoretically and we expect them to be well measured
experimentally \cite{Hrynova} in the near future.

 From the effective Hamiltonian at the quark level for $B$ decays
\cite{Hamiltonian}, the amplitudes of $B^-\to p\bar p K^-$ and
$B^-\to p\bar p K^{*-}$ are approximately given by
\cite{ChuaHouTsai2,HY,Geng3,Tsai,AngdisppK,ppKstar}
\begin{eqnarray}\label{eq1}
{\cal A}_K&\simeq&i\frac{G_F}{\sqrt 2}m_b f_K\bigg[\alpha_K\langle
p\bar p|\bar u b|B^-\rangle+\beta_K\langle p\bar p|\bar u\gamma_5
b|B^-\rangle\bigg]\,, \nonumber\\
{\cal A}_{K^*}&\simeq&\frac{G_F}{\sqrt
2}m_{K^*}f_{K^*}\varepsilon^{\mu}\alpha_{K^*}\langle p\bar p|\bar
u\gamma_\mu(1-\gamma_5) b|B^-\rangle\;,
\end{eqnarray}
respectively,
where $G_F$ is the Fermi constant,
 $f_{K^{(*)}}$ is the meson decay constant, given by
$\langle K^-|\bar s\gamma_\mu \gamma_5 u|0\rangle=-if_K q_\mu$
($\langle K^{*-}|\bar s\gamma_\mu u|0\rangle
=m_{K^*} f_{K^*}\varepsilon_\mu$)
with $q_\mu$ ($\varepsilon_\mu$) being the four momentum (polarization) of $K^-$ ($K^{*-}$), and $\alpha_{K^{(*)}}$ and $\beta_{K}$ are defined by
\begin{eqnarray}\label{eq2}
\alpha_K&\equiv& V_{ub}V_{us}^*a_1-V_{tb}V_{ts}^*\bigg[a_4+ a_6\frac{2 m_K^2}{m_b m_s}\bigg]\;,\nonumber\\
\beta_K&\equiv& V_{ub}V_{us}^*a_1-V_{tb}V_{ts}^*\bigg[a_4- a_6\frac{2 m_K^2}{m_b m_s}\bigg]\;,\nonumber\\
\alpha_{K^*}&\equiv& V_{ub}V_{us}^*a_1-V_{tb}V_{ts}^*a_4\;,
\end{eqnarray}
where $V_{ij}$ are the CKM matrix elements
and $a_i$ ($i=1,4,6$) are given by
\begin{eqnarray}\label{a146}
a_1=c_1^{eff}+\frac{1}{N_c}c_2^{eff}\;,\;a_4=c_4^{eff}+\frac{1}{N_c}c_3^{eff}\;,\;a_6=c_6^{eff}+\frac{1}{N_c}c_5^{eff}\;,
\end{eqnarray}
with $c_i^{eff}\;(i=1,2, ..., 6)$ being effective Wilson
coefficients (WC's) shown in Refs. \cite{Hamiltonian} and $N_c$
the  color number for the color-octet terms.  We note that
for the decay amplitudes in Eq. (\ref{eq1}) we have neglected
the small contributions \cite{AngdisppK,ppKstar} from
$\langle
p\bar p|J_1|0\rangle\langle K^{(*)}|J_2|B\rangle $  
involving the 
 $vacuum \to p\bar p$  time-like  
baryonic form factors \cite{NF}, where
$J_{1,2}$ can be (axial-)vector or (pseudo)scalar currents.
However, 
in our numerical analysis
we will keep all amplitudes including the ones neglected in 
Eq. (\ref{eq1}).

The direct CP asymmetries in
$B^{\pm}\to p\bar p M^{\pm}$ with $M=K$ and $K^{*}$
are defined by
\begin{eqnarray}\label{Acp}
A_{CP}(M)=\frac{\Gamma(B^-\to p\bar p
M^-)-\Gamma(B^+\to p\bar p M^+)}{\Gamma(B^-\to p\bar p M^-)
+\Gamma(B^+\to
p\bar p M^+)}\,,
\end{eqnarray}
while the decay rates can be evaluated from Eq. (\ref{eq1}) after integrations over  the three-body phase spaces.
Interestingly  from Eqs. (\ref{eq1}) and (\ref{Acp})
we derive the simple results:
\begin{eqnarray}\label{Acp2}
A_{CP}(K)&=\bigg(&\frac{|\alpha_{K}|^2-|\bar \alpha_{K}|^2}{|\alpha_{K}|^2+|\bar \alpha_{K}|^2}+R_{\beta}^-\bigg)/(1+R_{\beta}^+)\,,\nonumber\\
A_{CP}(K^*)&=&\frac{|\alpha_{K^{*}}|^2-|\bar
\alpha_{K^{*}}|^2}{|\alpha_{K^{*}}|^2+|\bar \alpha_{K^{*}}|^2}\,,
\end{eqnarray}
where
\begin{eqnarray}\label{Rbeta}
R_{\beta}^{\pm}\equiv\bigg(\frac{|\beta_{K}|^2\pm|\bar\beta_{K}|^2}{|\alpha_{K}|^2+|\bar
\alpha_{K}|^2}\bigg)R
\end{eqnarray}
with $R$ being around $0.27$
and $\bar\alpha_{K^{(*)}}$ and $\bar\beta_{K^{(*)}}$ denote the
values of the corresponding antiparticles, respectively.
It is easy to see that
$A_{CP}(K^{*})$ is independent of the phase spaces as well as the
hadronic matrix elements. As a result, the hadron parts along with
their uncertainties in $A_{CP}(K^{*})$ are divided out in Eq.
(\ref{Acp2}). On the other hand, although $A_{CP}(K)$ seems to
suffer from the hadron uncertainty due to $R_{\beta}^{\pm}$ in Eq.
(\ref{Rbeta}), by noticing that ($R_{\beta}^{+}$,
$R_{\beta}^{-}$)$\simeq (0.047, 0.002)$ are small numbers,
as a rough estimate we may
neglect the $R_{\beta}^{\pm}$ terms in Eq.
(\ref{Rbeta}), which implies that the hadronic uncertainty 
for $A_{CP}(K)$ is also suppressed. 
We note that the CP asymmetries in Eq. (\ref{Rbeta}) are related to
the weak phase of $\gamma (\phi_{3})$ \cite{pdg}.

In our numerical calculations, the CKM parameters are taken to be
\cite{pdg} $V^{\;}_{ub}V^{*}_{us}=A\lambda^4(\rho-i\eta)$ and
$V^{\;}_{tb}V^{*}_{ts}=-A\lambda^2$ with $A=0.818$,
$\lambda=0.2272$, the values of $(\rho,\eta)$ are $(0.221,0.340)$
\cite{pdg}. We remark that $a_i$  contain both weak and strong
phases, induced by $\eta$ and quark-loop rescatterings
\cite{Wolfenstein}, respectively. Explicitly, at the scale $m_b$
and $N_c$=3, we obtain a set of $a_1$, $a_4$, and $a_6$ as
follows:
\begin{eqnarray}\label{set1}
a_1&=&1.05\;,\nonumber\\
a_4&=&\big[(-427.8\mp 9.1\eta-3.9\rho)+i(-83.2\pm 3.9\eta-9.1\rho)\big]\times
10^{-4} \;,\nonumber\\
a_6&=&\big[(-595.5\mp 9.1\eta-3.9\rho)+i(-83.2\pm 3.9\eta-9.1\rho)\big]\times
10^{-4}\;,
\end{eqnarray}
for the $b\to s$ ($\bar b\to \bar s$) transition.
Our results on the direct CP violation are shown
in Table \ref{pre}.
\begin{table}[h]
\caption{Direct CP asymmetries in $B^\pm\to p\bar p M^{\pm}\ (M=K,K^{*})$,
where the errors in our work represent the possible fluctuations induced from non-factorizable effects, time-like
baryonic form factors and CKM matrix elements, respectively.
}\label{pre}
\begin{tabular}{|c||c|c|}
\hline 
$A_{CP}(M)$
&$A_{CP}(K)$&$A_{CP}( K^{*})$\\\hline
our work&$0.06\pm 0.01\pm 0.003\pm 0.01$&$0.22^{+0.04}_{-0.03}\pm 0.01\pm 0.01$\\
BELLE \cite{Wang1}&$-0.05\pm0.11\pm0.01$          &--------\\
BABAR \cite{CPppK,Hrynova}&$-0.13^{+0.07}_{-0.08}\pm0.04$ &$0.26\pm 0.19$\\\hline
\end{tabular}
\end{table}
In the table, we have included the possible
fluctuations induced from non-factorizable effects, time-like
baryonic form factors and CKM matrix elements, respectively. 
We note that 
the uncertainties from time-like baryonic form factors  are constrained by the data of $\bar B^0\to
n\bar p D^{*+}$ and $\bar B^0\to\Lambda \bar p \pi^+$ \cite{NF}
and the errors on the CKM elements are from $\rho$ and $\eta$
 given in Ref.
\cite{pdg}.
It is interesting to point out that
the large value of $A_{CP}(B^\pm\to p\bar p
K^{*\pm})$=22\% is in agreement with  the BABAR data of $(26\pm 19)\%$ as given in Ref.
\cite{Hrynova}. However, taken at face value; the sign of our prediction  $A_{CP}(K)\sim 0.06$
for $B^{\pm}\to p\bar p K^\pm$ is different from those
of $-0.05\pm0.11\pm 0.01$ and
$-0.13^{+0.07}_{-0.08}\pm0.04$ measured by the BELLE \cite{Wang1}
and BABAR \cite{CPppK,Hrynova} Collaborations,
respectively. Since the uncertainties of both experiments are still large
it is too early to make a firm conclusion.

In term of  the hadronization approach in
Eqs. (\ref{eq1}) and (\ref{eq2}), there usually exist some
uncertainties. These come  from nonfactorizable effects when
gluons are attached to all hadrons, annihilation contributions when the $B$ meson
decays into the vacuum by the W boson emitting or exchange as well as
final state interactions. We find that they are all small in
our case as shown in Table \ref{pre}. Our  reasons may be outlined as follows:
\begin{enumerate}
\item
 Although the
nonfactorizable terms cannot be directly and unambiguiously figured out by
theoretical calculations, in the generalized factorization method
\cite{Hamiltonian} we could estimate the uncertainty by
parameterizing $N_c$ in Eq. (\ref{a146}) as  the effective color
number $N_c^{eff}$ running from $2$ to $\infty$.
 Explicitly using this  we find that
the maximal deviations for $A_{CP}(K^{(*)})$ are less than $0.01$ and
$0.04$, respectively.
\item In the perturbative QCD approach (PQCD), the
threshold effects measured by the experiments can be explained by
 power expanding the form factors in terms of $1/t^{n}$, where $t\equiv
(p_p+p_{\bar p})^2$ and $n\geq 2$ due to gluon propagators
attaching to valence quarks for the proton-antiproton pair
\cite{ChuaHouTsai2,AngdisppK,Brodsky1,Brodsky2,Brodsky3}. The
amplitudes from the annihilation terms are then  much suppressed
as $t\to m^{2}_{b}$ which is the squared transmitting energy.
\item For the
final state interactions, the most possible source is via the
two-particle rescattering to the proton pair, such as $B\to M_1 M_2
K^{(*)}\to p\bar p K^{(*)}$ with $M_{1,2}$ representing the
 meson states. However, such processes would shape the curve associated
with the phase spaces in the decay rate distributions
\cite{LambdaCBELLE,LambdaCChen,LambdaCCheng}, which have been
excluded in the charmless baryonic $B$ decay experiments. Other
types of the final state interactions have been parameterized into
the $B^-\to p\bar p$ form factors with the data in Ref.
\cite{angularppK}, which are insensitive to $A_{CP}(K^{(*)})$  as seen in
Eq. (\ref{Acp2}).
\end{enumerate}

In summary,  we have shown that the $B^\pm\to p\bar p K^{(*)\pm}$
decays can be important tools to study the direct CP violation in
the charged $B$ system and we found that the  theoretical
hadronic uncertainties are small.
 In particular, we have found
that $A_{CP}(B^\pm\to p\bar p K^{*\pm})\simeq 22\%$ is large and
this is supported by the BABAR data \cite{Hrynova} of $(26\pm
19)\%$, whereas our prediction for $B^\pm\to p\bar p K^{\pm}$ has
a different sign in comparison with the BELLE \cite{Wang1} and
BABAR  \cite{CPppK,Hrynova} data. More precise measurements are
 needed in the $B$ experiments at the current and future
B-factories. 
It is
clear that these direct CP asymmetries should be used to test the SM and search for new physics.
Finally, we note that our study on the direct CP
violation can be extended to other modes, such as $B^{\pm}\to
p\bar{p}(\pi^{\pm},\rho^{\pm})$ and $B^{\pm}\to
\Lambda\bar{\Lambda}K^{(*)\pm}$. Explicitly, we predict that
$A_{CP}\sim -6\%,\ -3\%,\ 3\%$ and $-1\%$ for the above decays, respectively.\\

This work is supported in part by the National Science Council of
R.O.C. under Grant \#s:
NSC-94-2112-M-007-(004,005) and NSC-95-2112-M-007-059-MY3
and the Natural Science and Engineering Council of Canada.


\begin{thebibliography}{99}
\bibitem{CKM} N. Cabibbo, Phys. Rev. Lett. {\bf 10}, 531 (1963);
M. Kobayashi and K. Maskawa, Prog. Theor. Phys. {\bf 49}, 652
(1973).
\bibitem{JPsiAcpBABAR}
B.~Aubert {\it et al.} [BABAR Collaboration], Phys.\ Rev. \ Lett. {\bf 89}, 201802 (2002).
\bibitem{JPsiAcpBELLE}
K.~Abe {\it et al.}  [BELLE Collaboration], Phys.\ Rev.\ D {\bf 66}, 071102 (2002).

\bibitem{KpiAcpBABAR} B.~Aubert {\it et al.},[BABAR Collaboration], Phys. Rev. Lett. {\bf 93}, 131801 (2004).
\bibitem{KpiAcpBELLE} Y.~Chao {\it et al.}, [BELLE Collaboration], Phys.\ Rev.\ Lett.  {\bf 93}, 191802 (2004).
\bibitem{pipiAcpBELLE}
K.~Abe {\it et al.} [BELLE Collaboration], arXiv:hep-ex/0608035;
Phys.\ Rev.\ Lett.\  {\bf 95}, 101801 (2005).
\bibitem{charged} 
B.~Aubert {\it et al.}  [BABAR Collaboration], Phys.\ Rev.\ Lett.\  {\bf 94}, 181802 (2005).

\bibitem{Buras1} 
A.~J.~Buras, R.~Fleischer, S.~Recksiegel and F.~Schwab, Phys.\ Rev.\ Lett.\  {\bf 92}, 101804 (2004).
\bibitem{Buras2} 
A.~J.~Buras, R.~Fleischer, S.~Recksiegel and F.~Schwab, Nucl.\ Phys.\ B {\bf 697}, 133 (2004).

\bibitem{Botella}
F.~J.~Botella, G.~C.~Branco, M.~Nebot and M.~N.~Rebelo, Nucl.\ Phys.\ B {\bf 725}, 155 (2005).

\bibitem{Ali:2004dq}
  A.~Ali,
  Int.\ J.\ Mod.\ Phys.\ A {\bf 20} (2005) 5080
  [arXiv:hep-ph/0412128].

\bibitem{Brev}
 H.~n.~Li,
  arXiv:hep-ph/0605331; and references therein.

  \bibitem{Gronau:2006xu}
  M.~Gronau and J.~L.~Rosner,
  arXiv:hep-ph/0608040.

  \bibitem{Lipkin:2006pc}
  H.~J.~Lipkin,
  arXiv:hep-ph/0608284.

 \bibitem{ICHEP06-1}
 E. DiMarco, talk presented at ICHEP06, Moscow, Russia, July 26-August 2, 2006.

  \bibitem{ICHEP06-2}
 Y. Unno, talk presented at ICHEP06, Moscow, Russia, July 26-August 2, 2006.

\bibitem{Wang1} M.Z. Wang {\it et al.} [BELLE Collaboration], Phys. Rev. Lett. {\bf92}, 131801 (2004).
\bibitem{angularppK}
M.~Z.~Wang {\it et al.} [BELLE Collaboration], Phys.\ Lett. {\bf B617}, 141 (2005).
\bibitem{CPppK}
B.~Aubert {\it et al.} [BABAR Collaboration], Phys.\ Rev. {\bf D72}, 051101 (2005).

\bibitem{Threshold}
W.S.~Hou and A.~Soni, Phys. Rev. Lett.  {\bf 86}, 4247 (2001).

\bibitem{Hrynova} T.B. Hryn'ova, ``Study of B Meson Decays to $p\bar{p}h$ Final
States'', Ph.D. thesis, Stanford University (2006).
\bibitem{Hamiltonian}
Y.~H.~Chen, H.~Y.~Cheng, B.~Tseng and K.~C.~Yang, Phys.\ Rev. {\bf
D60}, 094014 (1999);
H.~Y.~Cheng and K.~C.~Yang, Phys.\ Rev. {\bf D62}, 054029 (2000).





\bibitem{ChuaHouTsai2}
C.~K.~Chua, W.~S.~Hou and S.~Y.~Tsai, Phys.\ Rev. {\bf D66}, 054004 (2002).
\bibitem{HY}
H.~Y.~Cheng and K.~C.~Yang, Phys.\ Rev. {\bf D66}, 014020 (2002).



\bibitem{Geng3}
C.~Q.~Geng and Y.~K.~Hsiao, Phys.\ Lett. {\bf B619}, 305 (2005).
\bibitem{Tsai} S. Y. Tsai, ``Study of Three-body Baryonic B
Decays'', Ph. D thesis, National Taiwan University (2005).
\bibitem{AngdisppK}
C.Q. Geng and Y.K. Hsiao, arXiv:hep-ph/0606141.
\bibitem{ppKstar}
C.~Q.~Geng, Y.~K.~Hsiao and J.~N.~Ng, in preparation.
\bibitem{NF} C.Q. Geng and Y.K. Hsiao, hep-ph/0606036.


\bibitem{pdg} W.~M.~Yao {\it et al.}  [Particle Data Group], J.\ Phys.\ G {\bf 33}, 1 (2006).

\bibitem{Wolfenstein} 
L.~Wolfenstein, Phys.\ Rev.\ D {\bf 43}, 151 (1991).

\bibitem{Brodsky1}
G.~P.~Lepage and S.~J.~Brodsky, Phys.\ Rev.\ Lett.\  {\bf 43},
545(1979) [Erratum-ibid.\  {\bf 43}, 1625 (1979)].
\bibitem{Brodsky2}
G.~P.~Lepage and S.~J.~Brodsky, Phys.\ Rev.\ {\bf D22}, 2157 (1980).
\bibitem{Brodsky3}
S.~J.~Brodsky, G.~P.~Lepage and S.~A.~A.~Zaidi, Phys.\ Rev.\ {\bf D23}, 1152 (1981).









\bibitem{LambdaCChen} 
C.~H.~Chen, Phys.\ Lett.\ B {\bf 638}, 214 (2006).
\bibitem{LambdaCBELLE} 
K.~Abe {\it et al.}  [BELLE Collaboration], arXiv:hep-ex/0508015.
\bibitem{LambdaCCheng} 
H.~Y.~Cheng, C.~K.~Chua and S.~Y.~Tsai, Phys.\ Rev.\ D {\bf 73}, 074015 (2006).

\end{thebibliography}
\end{document}